# The origin of mechanical enhancement in polymer nanoparticle composites with ultra-high nanoparticle loading

Emily Y. Lin[1], Amalie L. Frischknecht[2], Robert A. Riggleman[1]

1. Department of Chemical and Biomolecular Engineering, University of Pennsylvania, Philadelphia, PA 19104 USA
2. Center for Integrated Nanotechnologies, Sandia National Laboratories, Albuquerque, NM 87185 USA

# 1 Abstract

Polymer nanoparticle composites (PNC) with ultra high loading of nanoparticles (> 50%) have been shown to exhibit markedly improved strength, stiffness, and toughness simultaneously compared to the neat systems of their components. Recent experimental studies on the effect of polymer fill fraction in these highly loaded PNCs reveal that even at low polymer fill fractions, hardness and modulus increase significantly. In this work, we aim to understand the origin of these performance enhancements by examining the dynamics of both polymer and nanoparticles (NP) under tensile deformation. We perform molecular dynamics (MD) simulations of coarse-grained, glassy polymer in random-close-packed nanoparticle packings with a varying polymer fill fraction. We characterize the mechanical properties of the PNC systems, compare the NP rearrangement behavior, and study the polymer segmental and chain-level dynamics during deformation below the polymer glass transition. Our simulation results confirm the experimentally-observed increase in modulus at low polymer fill fractions, and we provide evidence that the source of mechanical enhancement is the polymer bridging effect.

# 2 Introduction

Random-close-packed nanoparticle (NP) films have a wide range of potential applications.[1–9] However, many applications are limited due to the brittle nature of the NP films.[10–12] The NPs interact with each other via relatively weak van der Waals interactions, and even slight mechanical perturbations can cause avalanche-like rearrangements that ultimately lead to shear-banding and fracture.[13–16] Numerous biological systems, such as human teeth, contain microstructures that overcome this brittle failure by combining inorganic particles with organic polymers to form composite materials.[17] In the past decade, there have been numerous advances in fabricating biomimetic structures that take advantage of the toughening of densely packed nanoparticles with the addition of polymer.[18,19] Fabricating polymer nanoparticle composites (PNC) with ultra high, near random close packed NP loading has been challenging with standard fabrication techniques. Recently, this challenge was overcome by infiltrating polymer into the NP packing using thermal annealing using capillary rise infiltration (CaRI),[20] solvent annealing,[21] or by leaching of mobile species in an elastomeric network.[22] Depending on the polymer-to-NP ratio, the polymer fill fraction in the final PNC can be tuned, resulting in undersaturated PNCs with tunable porosity at high NP loadings.

Additionally, it has been demonstrated that these PNCs with even a low polymer fill fraction have dramatic enhancements in mechanical properties compared to the neat NP films.[23,24] From these experiments, it was hypothesized that polymer bridging between nanoparticles may be the cause of the significantly increased fracture toughness of the PNC compared to the neat NP packing.[24]

Liquid bridging is well known to strengthen granular materials through capillary forces holding the particles together.[25–34] What remains poorly understood is if macromolecular liquids, such as polymers, can also lead to bridging and how the effect is changed when the constituent particles are nanoscopic in size. For a glassy PNC with low NP loading, Genix et al. recently showed that the mechanical properties can be tuned by changing the molecular weight of the polymer component.[35] Because of the low NP fill fractions, low molecular weight polymers cannot bridge multiple NPs and cause an increase in cohesive strength. When higher molecular weight polymers are used, a single polymer chain can contact more than one NP, and the addition of connectivity between polymer monomers lends an additional force that enhances the stiffness of the PNCs. In PNCs where the NP loading approaches random-close-packing and both liquid bridging and polymer bridging are present, the enhancement brought on by increasing molecular weight is not observed.[24] This suggests an antagonistic effect which adds to the mystery behind PNC's improvements in mechanical properties.

Another distinction from simple liquid bridging is that the polymeric component is often in a glassy state rather than a liquid state when the deformation occurs. Since polymer glass mechanics depend on many parameters such as temperature and sample preparation, the resulting composites may have different mechanical properties depending on their processing history.[36–47] The viscoelastic processes in the glassy polymer bridges between NPs are very different from the viscous dissipation experienced in liquid bridges. Monomeric level dynamics, such as segmental mobility, have been linked to macroscopic mechanical behavior due to deformation-induced mobility that allows polymer glasses to flow.[43–52] In inhomogeneous systems, segmental mobility has been shown to deviate drastically from bulk-like behavior,[50,51,53–57] consequently, it has become an important open question. Although several studies in the past decade focused on mechanical reinforcement in PNCs with low fill fractions of NPs,[52,58–65] to the best of our knowledge, there have been no prior microscopic studies that use non-equilibrium molecular dynamics simulations to thoroughly examine glassy polymer bridging of NPs in a densely packed system such as PNCs generated using CaRI or UCaRI. The molecular mechanism underlying the improvement in mechanical properties of polymer infiltrated nanoparticle films (PINFs) with high nanoparticle loading remains elusive.

In this study, we investigate the origin of the enhancement in mechanical properties in PINFs that resemble those formed using CaRI, and we test whether polymer bridging is the mechanism responsible for these enhancements.[23,24] We observe an increase in elastic modulus, yield stress, and yield strain in PINFs with increasing polymer fill fraction, similar to the increases found in prior experimental results. We find that a high fraction of polymer chains connect multiple nanoparticles, which allows the polymers to better accommodate strain without breaking, in contrast to simple liquids. Finally, the polymer chains in PINFs show faster-than-bulk segmental dynamics during deformation below $T_g$, and the polymer chains exhibit heterogeneous, bimodal segmental dynamics; the bridging chains have slower rearrangements than the chains that do not contact nanoparticles. Taken together our results provide a molecular picture of mechanical reinforcement and dissipation mechanisms in PINFs.

## 3 Methods

All simulations in this study are performed using the molecular dynamics (MD) simulation package LAMMPS[66, 67] with periodic boundary conditions in all dimensions. All non-bonded interactions are described using standard 12-6 Lennard Jones (LJ) pair potentials with a cutoff radius of $1.75\sigma$: $U_{ij}^{nb} = 4\epsilon \left[ \left( \frac{\sigma}{r_{ij}} \right)^{12} - \left( \frac{\sigma}{r_{ij}} \right)^{6} \right] - U_{rcut}^{nb}$ All quantities reported here are in reduced LJ units: temperature $T = kT^*/\epsilon$, and time $\tau_{LJ} = t^*\sqrt{\epsilon/m\sigma^2}$, where $m$ represents the mass of a single LJ interaction site, and $T^*$ and $t^*$ represent temperature and time measured in laboratory units. For describing the bonded interactions, we used a harmonic bonding potential with a stiff spring constant: $U_{ij}^{b} = (k_h/2)(r_{ij} - \sigma)^2$, where $k_h = 2000\ \epsilon/\sigma^2$. This model is a variation on the canonical Kremer-Grest model.[68] We used coarse grained polymers with chain lengths of 50 bonded LJ monomers per chain, and our MD simulations are integrated with a simulation time step of $\delta t = 0.002\ \tau_{LJ}$. To determine the glass transition temperature ($T_g$) for the pure polymer system, three independent cooling simulations are used to locate the temperature at which the thermal expansion coefficient ($\alpha_T = (\delta v/\delta T)_p$) changes during cooling simulations. For the pure polymer, $T_g = 0.394 \pm 0.002$

Each NP is modeled as a hollow, rigid shell comprised of 1243 LJ interaction sites covering the surface, and a LJ site at the center. The surface LJ sites are at $r = 10$ from the center LJ site. The NPs each have the same effective diameter of $21\sigma$. For all of our PNC systems, we use a total of 110 NPs with varying amounts of polymer. The polymer fill fraction ($\phi$) is defined as the number density of polymer monomers in the void volume of the NP packing, i.e. $\phi = N_{poly}/(V_{box} - V_{NP})$, where $N_{poly}$ is the total number of polymer monomers, $V_{box}$ is the volume of the simulation box, and $V_{NP}$ is the total volume of all of the NPs.

Three independent configurations of PNCs for each $\phi$ are used to provide estimates of run-to-run variations. Each PNC is equilibrated at high temperature (T = 10.0) and low density ($\phi_{NP} = 19.4\%$) to ensure good mixing of polymer and NPs. Then the simulation box is shrunken to the desired final volume, which is calculated such that the NP volume fraction would be near the random close packed limit, and quenched to $T = 1.0$. The final NP fill fraction is $59.5\% \pm 0.1\%$. The high density PNCs are equilibrated at $T = 1.0$ until the simulation box size stabilizes before being rapidly quenched again to $T = 0.30$ ($T/T_g = 0.761$) at a cooling rate of $\dot{\Gamma} = \Delta T/\Delta t = 1 \times 10^{-3}$.

We deformed each PNC at constant temperature and constant true strain rate ($\dot{\epsilon} = 3 \times 10^{-5}$) by applying uniaxial tension, while constant pressure ($p = 0.25$) was maintained in the orthogonal directions. We strained each PNC in the x, y, and z directions independently, giving a total of nine deformations for each state point. The error bars in all figures are obtained by calculating the standard error of the nine samples for each $\phi$. The stress in the direction of deformation was monitored during the simulation, and used to determine the elastic modulus of the systems. Figure 1 shows a visualization of an example PNC at $\phi = 0.87$ before and after tensile deformation, visualized using Ovito.[69]

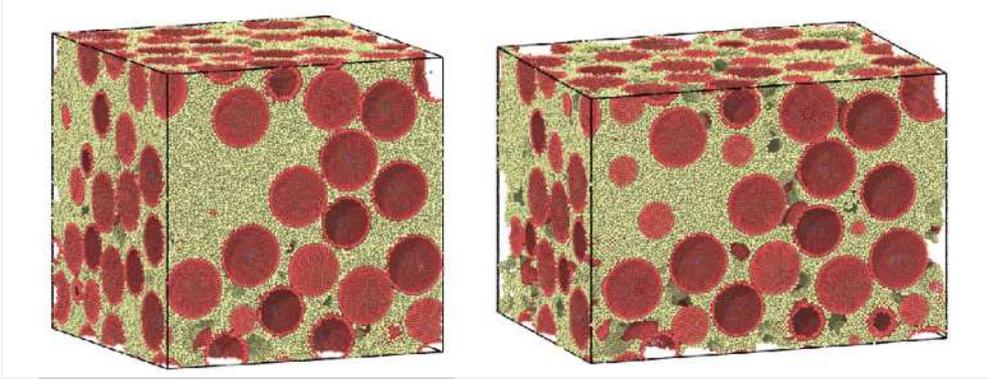

Figure 1: Visualization of PNC with $\phi = 0.87$ before (left) and after (right) tensile deformation. NPs are rigid hollow shells consisting of red monomers on the surfaces. Polymer chains are colored yellow.

Bond temporal auto-correlation functions, $C_b(t)$, are calculated to measure the dynamics of the glassy polymer chains during deformation. It has been shown previously that changes in mobility reported by the bond auto-correlation calculated during the deformation of glassy polymers agrees well with experimental trends.[47,49–51] $C_b(t)$ is defined as $C_b(t) = \left\langle P_2 \left[ \hat{b}(t) \cdot \hat{b}(0) \right] \right\rangle$, where $P_2$ is the second Legendre polynomial, and $\hat{b}$ is a unit vector aligned along the bonds. The affine contributions to the displacements are removed before calculating $C_b(t)$. A characteristic relaxation time is extracted by fitting $C_b(t)$ to the Kohlrausch-Williams-Watts (KWW) equation: $C_b(t) = C_o \exp\left(-t/\tau\right)^\beta$, where $C_o$, $\tau$, and $\beta$ are the fitting parameters, and $\tau$ is reported as the relaxation time.

Using polymer configurations sampled during the deformation, we calculated the local deviatoric strain rate ($J_2$) using a previously established method, without modifications to account for the 3D nature of our simulations.[70,71] $J_2$ is defined for particle $i$ as

$$J_{2_i}(\epsilon, \epsilon + \Delta\epsilon) = \frac{1}{\Delta\epsilon} \sqrt{\frac{1}{3} Tr \left[ \frac{1}{3}(\mathbf{J}_i^T \mathbf{J}_i - \mathbf{I}) - \frac{1}{3} Tr(\mathbf{J}_i^T \mathbf{J}_i - \mathbf{I}) \right]^2}$$

where $\mathbf{J_i}$ is the deformation gradient around particle i at strain $\epsilon$ after a lag strain of $\Delta\epsilon$ is applied, and $\mathbf{I}$ is the identity matrix. This local strain rate is calculated for each polymer monomer by calculating the best-fit local affine transformation matrix,[71] constructing the Lagrangian strain tensor, and removing the hydrostatic components. Particles with large $J_2$ values have a higher deviatoric strain rate in their local environment. The lag strain ($\Delta\epsilon$) we used for all $J_2$ calculations is 0.6%. The $J_2$ values are calculated throughout the deformation, and $J_2$ provides an additional measure of the monomers' mobility locally within the packing.

## 4 Results and discussion

We first compare the stress-strain behavior at different polymer fill fractions $\phi$ under tensile deformation. This comparison allows us to examine the effect of $\phi$ on the yield stress, yield strain, and elastic modulus of each system. Figure 2(a) shows the stress-strain curves for all of the $\phi$ studied, including the neat NP packing ($\phi = 0$), and Figure 2(b) shows the elastic moduli extracted as the initial slope of the stress-strain curves in the elastic regime. Below $\phi = 0.87$, as $\phi$ increases both the yield stress and the elastic modulus increase, which is consistent with experimental results.[23] Interestingly, the elastic modulus increase is non-monotonic with $\phi$; we observe a slight decrease at the largest value of $\phi$, which has not been observed in experiments. The origin of the decrease will be examined in more detail later, but one possible explanation for the difference between our simulation results and experiments could be the mode of deformation. Previous experiments used nano-indentation,[23,24] while we employ uniaxial tension.

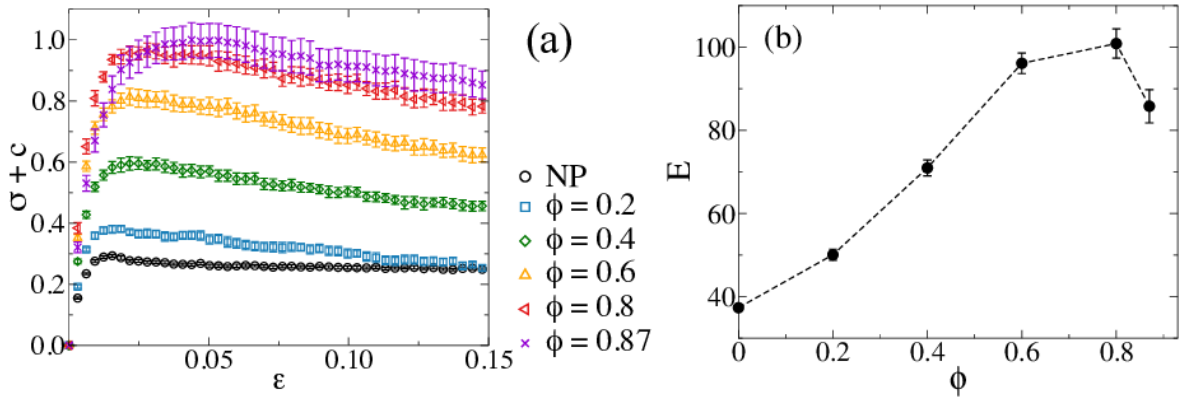

Figure 2: (a)Stress-strain curves for a neat NP packing and PNC systems. Stress values are offset by a constant, $c$, so that all curves start at the origin. Strain is calculated as engineering strain, using the simulation box dimensions. (b). Elastic modulus vs. polymer fill fraction, $\phi$, including neat NP packing $\phi = 0$.

To further elucidate the role of the polymer in the mechanical reinforcement of PNCs, we investigate the changes in NP contacts in the presence of the polymer. We calculate the nanoparticle-nanoparticle coordination number, $Z$, as the number of neighboring NPs within a cutoff distance of $2.5\sigma$, averaged for all NPs in each PNC system as a function of strain. Without deformation, the average coordination number, $\langle Z \rangle$, does not vary with time because the NPs are nearly jammed. The application of tensile strain pulls the NPs apart, therefore $\langle Z \rangle$ is expected to decrease as a function of strain. Figure 3 shows $\langle Z \rangle$ vs. strain for all $\phi$ studied. For the neat NP packing ($\phi = 0$), $\langle Z \rangle$ decreases rapidly with increasing strain, while for the PNCs, $\langle Z \rangle$ decreases much less rapidly. This behavior is consistent with our hypothesis that polymer chains in the PNC bridge the NPs together. NP contacts in the PNCs are maintained after strain because the polymer-NP interactions allow the polymer to form bridges between the NPs and "glue" the NPs together. An animated visualization of a polymer bridging NPs during deformation can be found in the supplemental material.

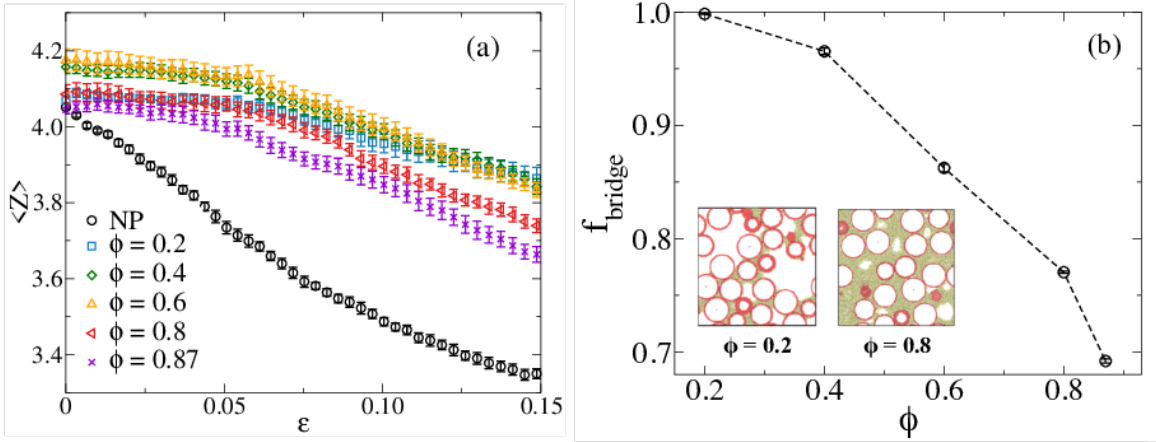

Figure 3: (a) Coordination number of the NP vs. applied strain for different fill fractions of polymer. (b) Average fraction of polymer chains that bridge between NPs ($f_{bridge}$) as a function of polymer fill fraction.

To characterize the presence of the polymer bridging phenomenon, we calculate the average fraction of polymer chains that make contact with more than one NP. This metric is defined as the number of polymer chains that contact more than one NP divided by the total number of polymer chains in the system, i.e., $f_{bridge} = N_{bridge}/N_{total}$, and a polymer chain is defined as in contact with a NP if any monomer is within the LJ interaction length scale (i.e., $r = 1.75$) of a NP. In Figure 3(b), we show $f_{bridge}$ as a function of $\phi$ before deformation, and we note $f_{bridge}$ for each $\phi$ does not change significantly during deformation. At $\phi = 0.2$, nearly all polymer chains make contacts with more than one NP (see the inset to Figure 3(b)). From $\phi = 0.2$ to $\phi = 0.4$, the number of polymer chains in the PNC doubles, while $f_{bridge}$ only decreases by a small amount, from 0.998 to 0.965. As $\phi$ further increases the decrease in $f_{bridge}$ is relatively slow, and the majority of polymer chains bridge NPs at all $\phi$, consistent with previous simulation results.[23] This slow decrease in $f_{bridge}$ indicates that polymer chains prefer to make bridges between particles due to polymer-NP interactions, and we expect that this effect would become even more pronounced with polymers of a higher chain length.

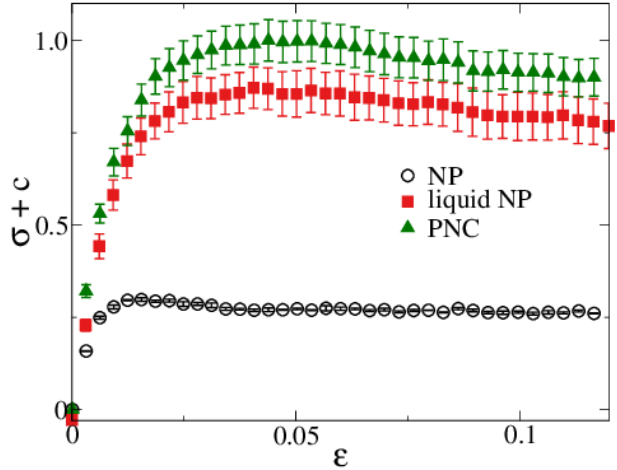

Figure 4: Stress-strain curves for the neat NP packing (black circles), liquid-filled NP packing (red squares), and polymer-filled NP packing (green triangles) at the same temperature, $T = 0.5$. The liquid or polymer fill fractions are $\phi = 0.87$.

To confirm that the presence of bonded polymers in the PNC further enhances mechanical properties compared to the liquid bridging in liquid-infiltrated NPs, we performed a set of simulations in which we raised the temperature to above the freezing point of the bulk LJ liquid, and subsequently deleted the bonds in the PNC system with the highest polymer fill fraction ($\phi = 0.87$) to create a liquid-NP system with the same NP configuration. We next deformed the resulting liquid-NP system to measure its stress/strain response. Figure 4 shows that although the liquid-NP system has a significantly higher yield stress and elastic modulus than the neat NP system, it has a lower yield stress and elastic modulus than the PNC system. This result provides additional support for our hypothesis that the presence of bonds in the PNC acts as an additional source of strength that further enhances the mechanical properties provided by simple liquid bridging, although a surprising amount of the increase in $E$ observed for the polymer-filled packings is captured when the packing is filled with a simple liquid.

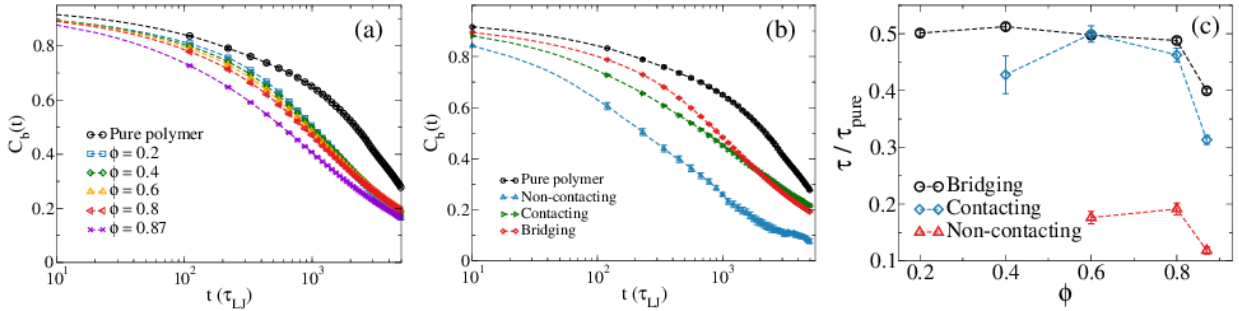

Figure 5: Bond autocorrelation decay during deformation. (a) $C_b(t)$ for pure polymer samples, and PNCs with all fill fractions. All polymer chains are considered in this subfigure. (b) Polymer chains are separated into three groups for $\phi = 0.6$ samples, and the $C_b(t)$ are calculated considering only chains in the given group. (c) Characteristic time scale of bond autocorrelation decay of bridging, contacting, and non-contacting polymer chains normalized by bulk time scale, as a function of polymer fill fraction.

To investigate how bridging affects polymer dynamics, we calculate the bond autocorrelation function, $C_b(t)$, during the deformation simulation, and $C_b(t)$ is plotted as a function of time for all PNCs in Figure 5(a). A faster decay of $C_b(t)$ indicates faster bond-reorientation, higher mobility of monomers, and increased dissipation of energy. We characterize $C_b(t)$ for three distinct groups of polymer chains: the "bridging" group of polymers, which contact at least two separate NPs, the "contacting" population of chains which contact only one NP, and the subset of polymers that make no contact with NPs ("Non-contacting"). We only present the non-contacting and contacting $C_b(t)$ data for high fill fractions where we have a sufficient number of chains in each group to obtain reliable estimates of $C_b(t)$. We also include $C_b(t)$ for the bulk polymer as a basis for comparison. In all cases, $C_b(t)$ decreases more rapidly in the packing than in a deformed bulk polymer at the same temperature, as shown in Figure 5(a). Presumably this is because more of the plastic deformation must be carried by a smaller volume of the sample, requiring the polymers to rearrange to accommodate any NP rearrangements as the deformation proceeds. As shown in Figure 5(b), the terminal decay behavior of both bridging and contacting groups is similar; however, bridging polymer chains show much slower decay initially, with a much more pronounced plateau. This means that the initial mobility of the monomers are slower when the entire chain is confined by two NPs. The non-contacting chains have $C_b(t)$ that decay much faster than contacting or bridging chains, indicating that these monomers have much higher mobility. These two observations hold true for all $\phi$, as indicated by the characteristic decay times in Figure 5(c). Interestingly, the non-contacting chains exhibit more mobility than those in bulk even at a large $\phi$ that approaches the density in the bulk ($\rho_{bulk} = 0.98$). This result implies that the NPs impart a significant change in polymer mobility during deformation that extends to chains that are not in contact with NPs.

Polymer dynamics are affected not only by contacts with the NPs, but also by the local strain rate, which is altered by the presence of the NP-imposed confinement. Since there are different degrees of confinement due to the near random-close-packed nature of the NPs, we define a different metric to categorize polymer monomers in terms of their proximity to NPs. We define $Z_{poly}$ to be the number of NPs a polymer monomer contacts (i.e., the distance between the monomer and the surface of NP is less than $2.5\sigma$), which allows us to quantify categorically the degree of confinement experienced by individual monomers. Because of the random-close-packed nature of the NPs, $Z_{poly}$ ranges from 0 to 4; however, only $Z_{poly}$ up to 3 is used due to the limited number of monomers with larger $Z_{poly}$ values. Using this definition, a $Z_{poly} = 0$ polymer monomer does not have any NP surface within $2.5\sigma$, which corresponds to the least confined scenario. Since the local geometry for the deforming system changes with respect to applied strain, the $Z_{poly}$ for each monomer is calculated as a function of strain. Next, we calculate the averaged polymer monomeric deviatoric strain rates, $J_2$, for each $Z_{poly}$ as a functions of strain, shown in Figure 6. $J_2$ is a useful metric to compare rearrangements in heterogeneous particle dynamics, because regions with high values of $J_2$ are correlated with the location of plastic rearrangements. For a low degree of confinement ($Z_{poly} = 0$), the $J_2$ of all the PNC systems are higher than that of the bulk system during the entire deformation (Figure 6). This suggests that the polymers in PNC systems experience more shear strain than those in the bulk, consistent with the faster dynamics shown in Figure 5. As $Z_{poly}$ increases, this trend slowly reverses. At the highest degree of confinement ($Z_{poly} =$ 3, Figure 6), the $J_2$ of nearly all the PNC systems are lower than the bulk values, suggesting the presence of slower monomer dynamics and less dissipation of plastic energy than in the pure polymer. The strong NP confinement effect on a monomeric level is consistent with the confinement effect on bond-reorientation, calculated using $C_b(t)$. Additionally, PNCs with lower $\phi$ consistently have higher $J_2$ values at all degrees of confinement, i.e., $Z_{poly}$ values. These observations can all be explained by the additional free volume that exists for lower $\phi$ and lower $Z_{poly}$. The confinement of polymer monomers plays a crucial role in changing the dynamics during deformation.

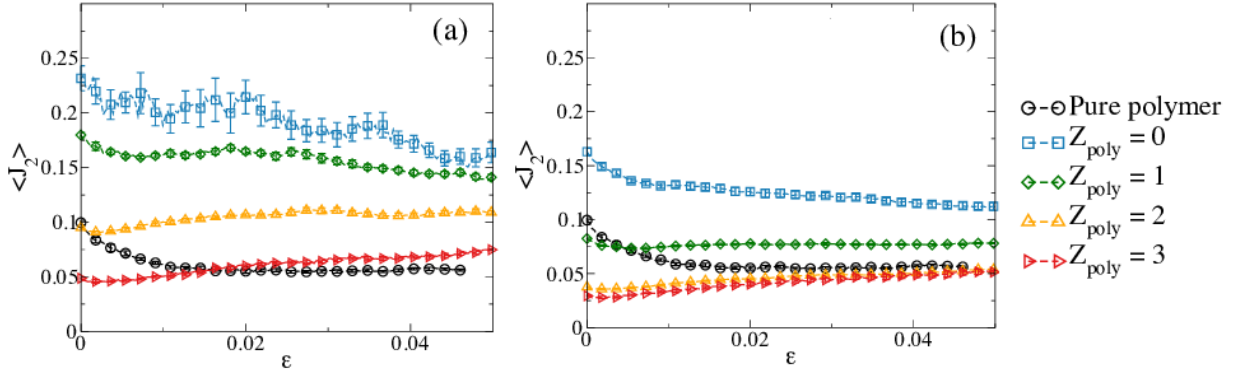

Figure 6: Local shear strain rates ($J_2$) vs. strain for different degree of confinement imposed by NPs: $Z_{poly}$ ranges from 0, least confined, to 3, most confined. (a) $\phi = 0.2$ PNC systems. (b). $\phi = 0.8$ PNC systems. Pure polymer samples are also included for comparison.

## 5 Summary

We used molecular dynamics (MD) simulations to show that in PNCs with ultra high NP loading, polymer bridging between NPs is a major contributor to the enhancement in mechanical properties that were observed in earlier experimental studies. We incorporated analyses from macroscopic and molecular-level perspectives to elucidate the polymer bridging phenomenon by examining the integrity of the NP packing, polymer monomer rearrangements, and polymer dynamics during deformation. We found that the polymer chains preferentially locate near NPs, and resist loss of NP contacts with increasing strain. We have also shown that the polymer dynamics in highly loaded PNC systems is heterogeneous, with bridging chains having much slower relaxation times than non-contacting chains. Interestingly, we found that the polymer dynamics in PNCs is faster than in the bulk, which may be explained by the increase in free volume introduced by the NPs. Future work on the polymer dynamics in CaRI PNCs with entangled polymer is underway to investigate any effect of entanglements on the mechanical properties and the effect of confinement on the entanglement network.

## 6 Acknowledgement

The authors would like to gratefully acknowledge our funding sources: National Science Foundation (NSF) PIRE grant 1545884. This work made use of computational resources provided through NSF Extreme Science and Engineering Discovery Environment (XSEDE) award TG-DMR150034. This work was performed, in part, at the Center for Integrated Nanotechnologies, an Office of Science User Facility operated for the U.S. Department of Energy (DOE) Office of Science. Sandia National Laboratories is a multimission laboratory managed and operated by National Technology & Engineering Solutions of Sandia, LLC, a wholly owned subsidiary of Honeywell International, Inc., for the U.S. DOE's National Nuclear Security Administration under contract DE-NA-0003525. The views expressed in the article do not necessarily represent the views of the U.S. DOE or the United States Government . We also thank Robert J. Ivancic, David Ring, Eric Bailey, Dr. Ryan Poling-Skutvik, Prof. Kevin Turner, Prof. Daeyeon Lee for helpful discussions.